\documentclass[aps,prd,superscriptaddress,nofootinbib,amsmath,amsfonts,preprintnumbers,notitlepage,10pt,english]{revtex4-1}
\usepackage{hyperref}
\usepackage{amsmath}
\usepackage{amssymb}
\usepackage{babel}
\usepackage{graphicx}
\usepackage{color}

\newcommand{\f}[2]{\frac{#1}{#2}}
\newcommand{\mk}[1]{\left( #1 \right)}
\newcommand{\kk}[1]{\left[ #1 \right]}

\newcommand{\be}{\begin{equation}}
\newcommand{\ee}{\end{equation}}
\def\detg{\sqrt{-g}}

\def\L{{\cal L}}

\def\K{{\cal K}}
\def\d{\delta}
\def\pd{\partial}
\def\Mpl{M_{\rm Pl}}
\def\tMpl{\tilde{M}_{\rm Pl}}
\newcommand{\es}[1]{e^{#1\sigma/\Mpl}}



\begin{document}

\preprint{}

\title{Stability of cosmological solutions in extended quasidilaton massive gravity}

\author{Hayato Motohashi}
\affiliation{Kavli Institute for Cosmological Physics, The University of Chicago, Chicago, Illinois 60637, U.S.A.}

\author{Wayne Hu}
\affiliation{Kavli Institute for Cosmological Physics, The University of Chicago, Chicago, Illinois 60637, U.S.A.}
\affiliation{Department of Astronomy \& Astrophysics, University of Chicago, Chicago IL 60637, U.S.A.}

\begin{abstract}
We consider the stability of self-accelerating solutions to extended quasidilaton massive gravity in the presence of matter.
By making a second  metric dynamical in this model, matter can 
cause it to evolve from a Lorentzian to Euclidean signature, triggering a ghost instability.
We study this possibility with scalar field matter as it can model a wide range of cosmological expansion 
histories.   For the $\Lambda$CDM expansion history, stability considerations substantially
limit the available parameter space while for a kinetic energy dominated expansion,
no choice of quasidilaton parameters is stable.  More generally these results show that
there is no mechanism intrinsic to the theory to forbid such pathologies from developing from stable initial
conditions and that stability can only be guaranteed for particular choices for the
matter configuration.
\end{abstract}

\date{\today}

\maketitle

\section{Introduction}

de Rham-Gabadadze-Tolley (dRGT) massive gravity~\cite{deRham:2010kj} is a theory with a massive graviton, which is constructed to remove the Boulware-Deser ghost.  In this theory, there are two metrics: the usual spacetime metric and a flat fiducial metric.   
It possesses a branch of self-accelerated solutions
\cite{Koyama:2011xz,Koyama:2011yg,Nieuwenhuizen:2011sq,Berezhiani:2011mt,Gratia:2012wt, Volkov:2012cf, Volkov:2012zb}
where the Universe undergoes  de Sitter expansion without a true cosmological constant.

However, because the fiducial metric is non-dynamical, the dRGT model breaks diffeomorphism invariance.  In the preferred unitary gauge coordinates where the fiducial metric is Minkowski, the spacetime metric does not take on the Friedmann-Lema\^{i}tre-Robertson-Walker (FLRW) form.  
Furthermore on the self-accelerating branch there is no
coordinate system where the two metrics are even simultaneously homogeneous and isotropic
for spatially flat or closed FLRW solutions \cite{D'Amico:2011jj}.  While
there exists open FLRW solutions where this is possible   \cite{Gumrukcuoglu:2011ew},
they are generally unstable to fluctuations \cite{Gumrukcuoglu:2011zh,DeFelice:2012mx}.
Though accelerating solutions where one of the two metrics are either inhomogeneous
or anisotropic do exist \cite{D'Amico:2011jj,Gratia:2012wt,Kobayashi:2012fz,Gumrukcuoglu:2012aa,Motohashi:2012jd}, this
feature of dRGT with a static flat fiducial metric is an obstacle in building a successful cosmology.

Many generalizations of the dRGT model focus on replacing the static flat fiducial
metric while retaining the Boulware-Deser ghost free form of the construction.
Quasidilaton massive gravity is 
one of such attempts
to make the fiducial metric dynamical.
Here the quasidilaton acts as a conformal rescaling of the fiducial metric and so
can accommodate the expansion of the Universe in both metrics \cite{D'Amico:2012zv}.   Unfortunately, in its original form the model suffers from ghost instabilities \cite{Gumrukcuoglu:2013nza,D'Amico:2013kya}.   
The extended quasidilaton model introduces an extra coupling term between the massive
graviton and quasidilaton that
cures this instability for vacuum self-accelerating solutions \cite{DeFelice:2013tsa}.  
However making the fiducial metric itself dynamical and dependent on the evolution of
the quasidilaton field, opens the possibility that in the presence of matter instabilities
develop.   It is the purpose of the present paper to investigate the stability of the extended quasidilaton massive gravity in cosmological solutions with matter.

This paper is organized as follows. 
In \S\ref{sec-eqmg}, we review the model and define notation. 
In \S\ref{sec-back}, we present homogeneous and isotropic background dynamics of the model. We show that it is still possible to have a self-accelerated solution in the presence of matter components.
In \S\ref{sec-pert}, we explore the scalar perturbations around the self-accelerated solution, and check their stability. Starting from summarizing the vacuum case, we then derive new conditions for stability with matter.
We conclude in \S\ref{sec-con}, and we provide techniques used in the main text in 
Appendix \ref{app-lcdm} and \ref{app-es}.

Throughout the paper, we will work in natural units where $c=1$, and the metric signature is $(-+++)$.

\section{Extended quasidilaton massive gravity with matter}
\label{sec-eqmg}

Extended quasidilaton massive gravity is defined by the action~\cite{DeFelice:2013tsa}
\be
S_g 
=\f{\Mpl^2}{2}\int d^4x \detg \kk{R+2m_g^2(\L_2+\alpha_3\L_3+\alpha_4\L_4) -\f{\omega}{\Mpl^2}\pd_\mu\sigma\pd^\mu\sigma},
\ee
where $m_g$ is the graviton mass, $\sigma$ is the quasidilaton scalar field, and $\omega$, $\alpha_3$, and $\alpha_4$ are dimensionless model parameters.
The graviton mass term is expressed by 
\begin{align}
	\L_2&=\f{1}{2}([\K]^2-[\K^2]),\nonumber\\
	\L_3&=\f{1}{6}([\K]^3-3[\K][\K^2]+2[\K^3]),\nonumber\\
	\L_4&=\f{1}{24}([\K]^4-6[\K]^2[\K^2]+3[\K^2]^2+8[\K][\K^3]-6[\K^4]).
\end{align}
Here, square brackets represent the trace of the enclosed matrix. 
The form of $\L_2$, $\L_3$, and $\L_4$ are the same as dRGT massive gravity but the matrix
$\K^\mu_{\;\,\nu}$ is given by
\be \K^\mu_{\;\,\nu}= \d^\mu_{\;\, \nu}-\es{} \mk{\sqrt{g^{-1}\tilde f}}^\mu_{\;\,\nu}, \ee
where $(g^{-1}\tilde f)^\mu_{\;\,\nu}=g^{\mu\rho}\tilde f_{\rho\nu}$ and $\sqrt{M}^\mu_{\;\,\nu}$ is understood as a root of the matrix: $\sqrt{M}^\mu_{\;\,\rho} \sqrt{M}^\rho_{\;\,\nu}= M^\mu_{\;\,\nu}$. 
There are two differences in $\K^\mu_{\;\,\nu}$ from dRGT massive gravity: the 
extended fiducial metric $\tilde f_{\mu\nu}$ which is disformally related to the fiducial metric $f_{\mu\nu}$
\begin{align}
\tilde f_{\mu\nu} &= f_{\mu\nu} - \f{\alpha_\sigma}{\Mpl^2m_g^2}\es{-2} \pd_\mu\sigma \pd_\nu\sigma,\nonumber\\
f_{\mu\nu} &= \eta_{ab} \pd_\mu\phi^a \pd_\nu\phi^b,
\label{eqn:fiducialmetrics}
\end{align}
and its coupling to the quasidilaton. 
Note that this disformal relation does not guarantee 
a Lorentzian signature to the extended fiducial metric.  More generally a disformal scaling which  does not
depend on the kinetic term of $\sigma$ itself does not by construction preserve the signature \cite{Bruneton:2007si}.   Thus we are interested in the question of whether matter can induce an evolution in $\sigma$ that changes the 
signature of this metric.

In (\ref{eqn:fiducialmetrics}) the 4  St\"{u}ckelberg fields  $\phi^a$ restore general covariance as they transform as spacetime scalars.  In addition, the form of the coupling is chosen so that under a transformation
\begin{equation}
\sigma \rightarrow \sigma+\sigma_0, \quad \phi^a \rightarrow e^{-\sigma_0/\Mpl} \phi^a,
\end{equation}
where $\sigma_0=$const.,
the extended fiducial metric transforms as
\begin{equation}
\tilde f_{\mu\nu}  \rightarrow e^{-2 \sigma_0/\Mpl} \tilde f_{\mu\nu},
\end{equation}
leaving the action invariant
as in the original quasidilaton model \cite{D'Amico:2012zv}.  The quasidilaton thus allows
a rescaling of the extended fiducial metric and in cosmological solutions plays a similar role to
the scale factor.
The coupling constant $\alpha_\sigma$ between the massive graviton and the quasidilaton is introduced to stabilize the self-accelerating  solution in the absence of matter \cite{DeFelice:2013tsa}.  Note that the extended fiducial metric is dynamical whereas $f_{\mu\nu}$ is
always a coordinatization of the standard Minkowski metric regardless of the dynamics of the
 St\"{u}ckelberg fields.

We are interested in how the background and the perturbations for the self-accelerating flat FLRW solution behave if we include matter component.  In order to consider a wide range
of cosmological background solutions, we take the matter to be a canonical scalar
field $\xi$ 
whose action is given by
\be
S_m 
=\int d^4x \detg \kk{-\f{1}{2}g^{\mu\nu} \pd_\mu\xi\pd_\nu\xi-V(\xi)}.
\ee
The total action $S=S_g+S_m$ is thus specified by 5 model parameters
$\{ m_g, \omega,\alpha_3,\alpha_4,\alpha_\sigma \}$ and a choice of the scalar field
potential $V(\xi)$.

\section{Background}
\label{sec-back}

The form of spatially flat
cosmological background solutions is defined by homogeneity and isotropy of the
spacetime and fiducial metrics, and quasidilaton and matter fields
\begin{align}
ds^2 &= -N(t)^2dt^2+a(t)^2\d_{ij}dx^idx^j, \nonumber\\
\phi^0&=\phi^0(t), \quad \phi^i=x^i,  \nonumber\\
\sigma&=\bar \sigma(t),\quad \xi=\bar \xi(t).
\end{align}
Here we have kept a general lapse function $N(t)$ so as to study its implied equation of motion before setting it to unity as in the conventional FLRW metric.
The extended fiducial metric is then given by
\begin{align}
-\tilde f_{00}&\equiv n(t)^2=(\dot \phi^0)^2+\f{\alpha_\sigma}{\Mpl^2m_g^2}\es{-2\bar}\dot{\bar\sigma}^2 , \nonumber \\
\tilde f_{ij}&=\d_{ij}.
\end{align}
It is convenient to introduce the following 
variables:
\be
H\equiv\f{\dot a}{Na}, \quad
X\equiv\f{e^{\bar\sigma/\Mpl}}{a},\quad
r\equiv\f{n}{N}a.
\ee
The Lagrangian at the background level is then given by
\begin{align}
\L=&\f{\Mpl^2}{2}a^3N\bigg[ 
6\mk{\f{\dot a^2}{N^2a^2}+\f{\ddot a}{N^2a}-\f{\dot N\dot a}{N^3a}}
 \notag\\
&+2m_g^2(X-1)[3(rX+X-2)-(X-1)(3rX+X-4)\alpha_3+(X-1)^2(rX-1)\alpha_4] \notag\\
&+\f{\omega \dot \sigma^2}{\Mpl^2N^2}+\f{1}{\Mpl^2}\mk{\f{\dot \xi^2}{N^2}-2V}
\bigg].
\end{align}

From this Lagrangian we can derive the equations of motion.  Variation of the
action with respect to $\phi^0$ gives
\be \f{d}{d t}\kk{\f{\dot \phi^0}{n}a^4 X(X-1)J}=0, \label{phi0} \ee
where 
\be J\equiv 3+3(1-X)\alpha_3+(1-X)^2\alpha_4. \ee
Since the St\"{u}ckelberg field do not couple to the matter field, this equation is the same in the presence or absence of matter and we follow Ref.~\cite{DeFelice:2013tsa} in studying its
solutions.
From (\ref{phi0}), we obtain $X(1-X)J\dot \phi^0/n\propto a^{-4}$ which asymptotically vanishes as the Universe expands. We focus on the branch with $J=0$, and hereafter $X$ shall denote the root of $J=0$. In this branch of cosmological solutions, 
$X\equiv e^{\bar\sigma/\Mpl}/a={\rm const}$ 
and  the quasidilaton in the background plays the same
role as the scale factor allowing the extended fiducial metric to scale with the expansion.
This solution implies
$\dot {\bar \sigma}=\Mpl N H$
and
\be \label{phi0sq} \mk{\f{\dot \phi^0}{n}}^2=1-\f{\alpha_\sigma \es{-2\bar}}{\Mpl^2m_g^2}\f{\dot{\bar \sigma}^2}{n^2}=1-\f{\alpha_\sigma H^2}{m_g^2X^2r^2}. \ee
If we insist that both the fiducial and extended fiducial metrics have a Lorentzian signature
then  $\dot \phi_0/n$ is real and
\be \label{conalpha} \alpha_\sigma<\f{m_g^2X^2r^2}{H^2}. \ee
For $r^2>0$, violation of this bound means that the fiducial metric loses its Lorentzian signature.

Variation with respect to  $N$ and $a$ give the Friedmann equations
\begin{align}
\label{fre1} 3\mk{1-\f{\omega}{6}}\Mpl^2 H^2 &= \Mpl^2 \Lambda_X + \f{\dot {\bar\xi}^2}{2}+V,\\
\label{fre2} -2\mk{1-\f{\omega}{6}}\Mpl^2 \dot H &= \dot {\bar\xi}^2.
\end{align}
After deriving the equation of motion for $N$, we set $N=1$ for the following. 
Here, we define
\be
\Lambda_X 
\equiv m_g^2(X-1)^2[(X-1)\alpha_3-3].
\ee
Therefore, the total energy consists of the matter component and an effective cosmological constant induced by the graviton mass term, which leads to a self-accelerated expansion of the Universe.   To make 
$\Lambda_X\sim m_g^2$ responsible for the late-time acceleration, one needs $m_g\sim H_0$ and its positivity requires
\be \label{cccon} (X-1)\alpha_3-3>0. \ee
In addition, we note that the effective gravitational constant for background is given by a
rescaling of the Planck mass
\begin{equation}
\tMpl^2 \equiv \Mpl^2 \mk{1-\f{\omega}{6}}
\end{equation}
and $\tMpl^2>0$ requires 
\be \label{conomega} \omega<6. \ee
By defining the effective critical density $\tilde \rho_{cr}\equiv 3 \tMpl^2 H_0^2$,
the Friedmann equations (\ref{fre1}), (\ref{fre2}) take their usual form.   In particular
for the $\Lambda$CDM expansion history with $\Omega_i\equiv \rho_i/ \tilde \rho_{cr}$,
$H^2/H_0^2=\Omega_\Lambda+\Omega_m a^{-3}$. 
Setting $\Omega_\Lambda$ to satisfy observational constrains determines  
$m_g/H_0$
as
\be \label{mgvalue} \f{m_g^2}{H_0^2}=\f{(6-\omega)\Omega_\Lambda}{2(X-1)^2[(X-1)\alpha_3-3]}. \ee

From the equation of motion for the quasidilaton $\bar \sigma$, we obtain
\be \label{req} r=1+\f{\omega(3 H^2+ \dot H)}{3m_g^2X^2[(X-1)\alpha_3-2]}. \ee
Therefore, $r$ is not constant in general, a crucial distinction from the case 
without a matter field.  It is only constant if  
$H$ itself is constant, or if $3H^2 + \dot H=0$.
In particular, the latter case implies that the Universe is dominated by the stiff matter, whose equation of state parameter is $w=1$. This phase could take place if the expansion is
dominated by the kinetic energy of the scalar field. 
In this case, $r=1$, which we shall see has interesting consequences for stability.

Finally, the matter field $\bar \xi$ obeys the usual equation for a minimally coupled scalar field
\be \label{eomxi} \ddot{\bar \xi} +3H\dot{\bar \xi} +\frac{d V}{d\xi}=0. \ee
Since the equation of state parameter for the scalar field is $w \ge -1$, it typically dominates
the energy density and the expansion rate in the past.   We shall use the flexibility in choosing
the potential to mimic the various stages of the standard $\Lambda$CDM model.  
In particular, we can reproduce  any power law expansion $a\propto t^p$ by using the potential for  power-law inflation. 
Furthermore,  it is possible to reproduce an expansion which is equivalent to that with nonrelativistic matter and a cosmological constant.
This case is studied in the Appendix \ref{app-lcdm}.

To summarize, we choose the model parameters, namely, $\{m_g, \omega, \alpha_3, \alpha_4, \alpha_\sigma\}$ in order to satisfy requirements on the background evolution.
Since
\be \label{xeq} 3+3(1-X)\alpha_3+(1-X)^2\alpha_4=0 \ee
on the self accelerating branch
and we also need to satisfy a condition on $X$ \eqref{cccon} for positivity of the effective cosmological constant, it is useful to choose first $\alpha_3$ and $X$ and determine $\alpha_4$ by \eqref{xeq}.
A specific example of a set of parameters which satisfy \eqref{xeq} and \eqref{cccon} is 
\be \label{exax} \alpha_3=4,\quad \alpha_4=9,\quad X=2.\ee
For $\omega$, we only need to satisfy \eqref{conomega} in order to  guarantee the 
positivity of the gravitational constant.
For $\alpha_\sigma$, 
\eqref{conalpha} is necessary if all metrics have Lorentzian signatures.
We shall see in the next section that this condition can be alternately viewed as a requirement for  the stability of fluctuations
around the background solution  which generalizes the vacuum results of Ref.~\cite{DeFelice:2013tsa}.   Then we set $m_g$ \eqref{mgvalue} using the observational data for $\Omega_\Lambda$.
For instance, for parameter set \eqref{exax}, $\omega=4$, and $\Omega_\Lambda=0.7$, we obtain $(m_g/H_0)^2=0.7$.
After specifying all the parameters, the evolution of $H(t)$ and $\bar \xi(t)$ are given by \eqref{fre2} and \eqref{eomxi}, and $r(t)$ is given by \eqref{req}.   Importantly, this makes the bound
on $\alpha_\sigma$ time dependent beyond the vacuum solutions.

\section{Scalar perturbations}
\label{sec-pert}

We will work in the unitary gauge, where the perturbation for the St\"{u}ckelberg field vanishes.
This gauge condition completely fixes the gauge degree of freedom and 
requires the most general parameterization of scalar metric fluctuations
\begin{align}
\d g_{00}&= -2\Phi,\nonumber\\
\d g_{0i}&= a\pd_i B,\nonumber\\
\d g_{ij}&= a^2\kk{2\d_{ij}\Psi+\mk{\pd_i\pd_j-\f{1}{3}\d_{ij}\pd_\ell\pd^\ell}E },
\end{align}
and two dimensionless perturbation for quasidilaton and matter field
\begin{align}
\sigma&=\bar\sigma+\Mpl\d\sigma,\nonumber\\
\xi&=\bar\xi+{\Mpl}\d\xi,
\end{align}
and we will work in a Fourier space.

Since the quadratic action does not have kinetic term for $B$ and $\Phi$ as expected, we can eliminate them by using their equations of motion.  We are then left with four variables, $\Psi$, $E$, $\d\sigma$, and $\d\xi$.

\subsection{Vacuum case}
\label{ssec-Lambda}

Let us first review the case without matter component.
To analyze this case, we only need to switch off $\xi$ and $V$, and use $\dot H=0$.
Then we have three variable $\Psi$, $E$, and $\d\sigma$. 
However, the kinetic terms for $\Psi$ and $\d \sigma$ can be combined in the form of $(\dot\Psi-\dot{\d\sigma})^2$.
Therefore, one nondynamical degree of freedom still remains in the quadratic Lagrangian. We define a new notation as 
\begin{align}
\phi_1&\equiv \Psi-\d\sigma \text{ : dynamical},\nonumber\\
\phi_2&\equiv E \text{ : dynamical},\nonumber\\
\phi_3&\equiv \Psi+\d\sigma \text{ : nondynamical}.
\end{align}
After integrating out $\phi_3$, the kinetic terms are $K_{ij} \dot \phi_i \dot \phi_j$ for $i$, $j=1,2$. The no-ghost condition is given by the positivity of all the eigenvalues of the kinematic matrix $K_{ij}$, which is equivalent to imposing
\begin{align}
\label{detKml} \det K&= \f{\Mpl^4 \omega^2 a^2 H^2 k^6}{r^2(r-1)^2} \f{2A(r-1)^2(k/aH)^2+3(\omega-6)(A-r^2)}{4(A-1)(k/aH)^2+\omega(6-\omega)} >0,\\
\label{K22ml} K_{22} &= \f{k^4\Mpl^2}{18} \f{\omega[2(A-1)(k/aH)^2+3(6-\omega)]}{4(A-1)(k/aH)^2+\omega(6-\omega)}>0,
\end{align} 
where 
\be A\equiv \f{\alpha_\sigma H^2}{m_g^2X^2}. \ee
Note that $H$ is given by \eqref{fre1} without the matter component, $r$ is given by \eqref{req} with $\dot H=0$, and both $H$ and $r$ are constant.

We would like to derive a condition for model parameters to make both  \eqref{detKml} and \eqref{K22ml} positive for all wavenumbers $k$.
We start from deriving necessary conditions from taking high-$k$ and low-$k$ limit.
For $k/aH \gg 1$, 
\be \f{A}{A-1}>0, \quad \omega>0. \ee
For $k/aH \ll 1$, the $K_{22}$ condition is automatically satisfied and
\be\f{r^2-A}{\omega}>0 , \ee
Therefore, in addition to $\omega<6$ from the positivity of the effective gravitational constant, the necessary condition for the stability is
\be \omega>0, \text{ and } A<r^2, \text{ and } [ A>1 \text{ or } A<0 ]. \ee 
Now let us check the sufficiency of the conditions. 
We note that $A<0$ is not sufficient. For instance, 
we can choose wavenumber
\be \mk{\f{k}{aH}}^2=\f{3(1-\epsilon)(6-\omega)}{2(1-A)}, \ee
which is positive by the virtue of $A<0$. 
Here, we choose some small positive $\epsilon$, which satisfies $0<\epsilon<{\rm Min} \{1, 6-\omega\}$.
For this wavenumber, $K_{22}$ is a positive number times
\be \f{\epsilon}{\omega-6+\epsilon}, \ee
which is negative.
On the other hand, $A>1$ is sufficient, because all the terms appeared in the expressions of $\det K$ and $K_{22}$ are positive for $A>1$, combined with $0<\omega <6$ and $A<r^2$.

Therefore, the no-ghost condition in the absence of matter component is given by
\be \label{ngLD} 0<\omega<6,\quad 1<\f{\alpha_\sigma H^2}{m_g^2X^2} <r^2. \ee
Note that we need $r>1$, namely, $(X-1)\alpha_3-2>0$, which is satisfied if we impose \eqref{cccon}. 
This condition is necessary to establish stability in asymptotic future of cosmological solutions of
the self accelerating branch.  Furthermore note that since $H$ and $r$ are constant here,
the stability condition for $\alpha_\sigma$  depends only on the choices for the other parameters of the quasidilaton model.
We shall next consider how these conditions generalize in the presence of matter.

\subsection{Matter with $-1 \le w < 1$}
\label{ssec-matter}

Now we turn our attention to 
examine the no-ghost condition in the presence of matter field, but we omit the case with $w=1$ for reasons which shall be made clear in \S\ref{ssec-stiff}.
In addition to the perturbation for the metric and the quasidilaton, we introduce matter perturbation $\d\xi$.
As in the absence of matter, the quadratic Lagrangian with matter does not have kinetic terms for $\Phi$ and $B$.
Thus, we can derive two constraint equations and make use of them to eliminate $\Phi$ and $B$.
After eliminating $\Phi$ and $B$, we are left with four perturbative variables, namely, $(E,\Psi,\d\sigma,\d\xi)$.
Without matter, we have two dynamical degrees of freedom. 
Therefore, we anticipate that with matter we should have three dynamical degrees of freedom, and there is one nondynamical degrees of freedom which should be expressed by certain linear combination of $(E,\Psi,\d\sigma,\d\xi)$.
Indeed, the determinant of the kinematic matrix for $(E,\Psi,\d\sigma,\d\xi)$ vanishes, which implies the existence of nondynamical field.
By examining the sub-kinematic matrices, we find that 
the kinematic matrix for $(\Psi, \d\sigma, \d\xi)$ is the minimal one whose determinant vanishes.
Let us denote the eigenvalues by $\lambda_1=0$, $\lambda_2$, $\lambda_3$, and their eigenvectors by $\vec v_1^T \equiv (1/\Xi,1/\Xi,1)$, $\vec v_2^T\equiv (v_{21}, v_{22}, 1)$, $\vec v_3^T\equiv (v_{31}, v_{32}, 1)$. Here, 
\begin{equation}
\Xi\equiv \frac{\dot{\bar\xi}}{\Mpl H}
\end{equation} and explicit forms for $\lambda_2$, $\lambda_3$, $\vec v_2$, and $\vec v_3$ are given in Appendix \ref{app-es}.
Now, we define a new basis thorough $(\Psi, \d\sigma, \d\xi)= (\vec v_1, \vec v_2, \vec v_3)(\psi_1,\psi_2,\psi_3)$, and diagonalize the sub-kinematic matrix.
After rewriting the quadratic Lagrangian in terms of new basis, we obtain the kinematic matrix for $\psi_1$, $\psi_2$, $\psi_3$, and $\psi_4\equiv E$ as
\be
\begin{pmatrix}
0 & 0 & 0 & 0\\
0 & \lambda_2 & 0 & * \\
0 & 0 & \lambda_3 & * \\
0 & * & * &  * 
\end{pmatrix},
\ee
where a star denotes nonzero components.
Therefore, we can eliminate the nondynamical variable $\psi_1$ by using its constraint equation, and end up with the quadratic Lagrangian for three dynamical degrees of freedom $(\psi_2, \psi_3, \psi_4)$.
We would like to examine the necessary conditions  for the kinematic matrix $K_{ij}$ for $i,j=2,3,4$
to be positive definite.
As in the matterless case, we investigate the sign of the determinants of the
matrix and its subblocks,  $\det K$, $K_{33} K_{44} - K_{34}^2$ and $K_{44}$ in the high-$k$ and low-$k$ limit.

First, let us focus on the high-$k$ limit. The leading order terms for $k/aH\gg 1$ are given by
\begin{align}
K_{44} &=\f{k^4}{72}\Mpl^2 a^3 (\omega+\Xi^2)+\cdots, \label{K44h} \\
\begin{vmatrix}
K_{33} & K_{34} \\
K_{34} & K_{44} \\
\end{vmatrix}
&= \f{\omega k^4}{144} \Mpl^4 a^6 (\Xi v_{32}-1 )^2+\cdots,  \label{detsubKh} \\
\det K &= \f{\omega^2 A k^2}{96 r^2 \Xi^2 (A-1)} \Mpl^6 a^{11} (3H^2+\dot H)(2+\Xi^2)^2[(1-2\omega)^2+2\Xi^2+\Xi^4] + \cdots. \label{detKh}
\end{align}
Here, $v_{32}$ is understood as the leading order term at the high-$k$ limit.
Let us determine the constraints on model parameters that are necessary for
these quantities to be positive. First, \eqref{K44h} is always positive. From \eqref{detsubKh} being positive, we have $\omega>0$. Since $\omega<6$ from the positivity of the effective gravitational constant, we obtain $0<\omega<6$. 
Last, \eqref{detKh} provides $A(A-1)>0$, namely, $(A>1~ \text{or}~ A<0)$. To also satisfy sufficient
conditions for the matterless case, we choose $A>1$.

Next, we focus on the low-$k$ limit. 
The leading order terms for $k/aH\ll 1$ are
\begin{align}
K_{44} &=\f{k^4}{12}\Mpl^2a^3 +\cdots, \label{K44l} \\
\begin{vmatrix}
K_{33} & K_{34} \\
K_{34} & K_{44} \\
\end{vmatrix}
&= \f{\omega (r^2 - A ) k^2}{8r^2 (r-1)^2} \Mpl^4 a^8 (3H^2+\dot H) (v_{31}-v_{32} )^2+\cdots , \label{detsubKl} \\
\det K &= \f{3\omega (r^2-A) k^2}{16 r^2 (r^2-1)} \Mpl^6 a^{11} H^2 [(v_{31}-v_{32})(\Xi v_{21}-1)-(v_{21}-v_{22})(\Xi v_{31}-1)]^2+\cdots. \label{detKl}
\end{align}
Here, $v_{21}$, $v_{22}$, $v_{31}$, $v_{32}$ are understood as the leading order terms at the low-$k$ limit.
From \eqref{detsubKl} and \eqref{detKl}, we obtain $A<r^2$. 

With the high-$k$ and low-$k$ results combined, the necessary conditions for stability are
\begin{equation} 
\label{ngmatter} 0<\omega<6,\quad \f{m_g^2X^2}{H^2(t)}<\alpha_\sigma<\f{m_g^2X^2}{H^2(t)}r^2(t), 
\end{equation}
which is identical to \eqref{ngLD} for the case without matter.
However, the crucial difference is that $H=H(t)$ and $r=r(t)$ are time dependent and the
condition must be satisfied for all time with a single value of the constant $\alpha_\sigma$.
This means that it is possible to choose parameters for which the system is initially
stable but evolve into an instability.   We shall show in the next section explicit examples that
do so.   Physically, this means that these backgrounds have their fiducial metrics evolve
from a Lorentzian to a Euclidean signature, thus triggering the instability.  By making the
second metric $\tilde f_{\mu\nu}$  dynamical in the extended quasidilaton scenario, stability depends not just on the intrinsic model parameters but also on the matter content and
evolution of the Universe.

Indeed, by introducing a scalar
field for the matter with an arbitrary potential, we have allowed for the possibility of
any expansion history for matter whose equation of state parameter varies from
$-1 \le w < 1$.   For instance, we can describe any expansion evolving as $w$=const. or
$a \propto t^{2/3(1+w)}$ by using the same
potential that describes power law inflation.
We shall next derive a more explicit condition from \eqref{ngmatter} with the help of the $\Lambda$CDM expansion.

\subsection{$\Lambda$CDM expansion history}
\label{ssec-lcdm}

Given the observational success of the $\Lambda$CDM expansion history, it
is worthwhile to explore the explicit constraints on parameters for this form.   
In  Appendix  \ref{app-lcdm}, we show that it is possible to construct two different 
 scalar field potentials that
reproduce the $\Lambda$CDM expansion history.   The first case is the usual
axion model where the field oscillates in a quadratic potential with $m \gg H$.  
The second, more novel case is a rolling field where the kinetic and potential energy are
driven to be equal through an attractor.   While these models have the same 
background expansion history, the axion model is equivalent to CDM in that
it is gravitationally unstable whereas the rolling field is not.    The two models
indicate that our condition (\ref{ngmatter}) is not dependent on whether matter
is gravitationally unstable in the linear regime.

Using the definition of $\Lambda_X$ and the expansion history $H^2 = H_0^2 ( \Omega_\Lambda+\Omega_m a^{-3} $), we obtain
\begin{align}
\label{ul1} \f{m_g^2X^2}{H^2}&=\frac{X^2}{2(X-1)^2}  \f{6-\omega}{(X-1)\alpha_3-3}\frac{\Omega_\Lambda}{\Omega_\Lambda+\Omega_m a^{-3} }, \\
\label{ul2} r&= 1+\f{\omega}{6-\omega}
\f{(X-1)^2}{X^2}
\f{(X-1)\alpha_3-3}{(X-1)\alpha_3-2}
\mk{2+\f{\Omega_m}{\Omega_\Lambda}a^{-3}}.
\end{align}
From \eqref{ul1} and \eqref{ul2}, the allowed range for $\alpha_\sigma$ \eqref{ngmatter} can be expressed in terms of $X$, $\alpha_3$, $\omega$, and $\Omega_\Lambda$.   Note that as $a$ decreases, the lower bound on $\alpha_\sigma$ monotonically weakens since $H\ge H_0$ while the upper bound asymptotically weakens since  $\lim_{a\rightarrow 0} r^2/H^2 \propto a^{-3}$.

\begin{figure}[t]
	\centering
	\includegraphics[width=75mm]{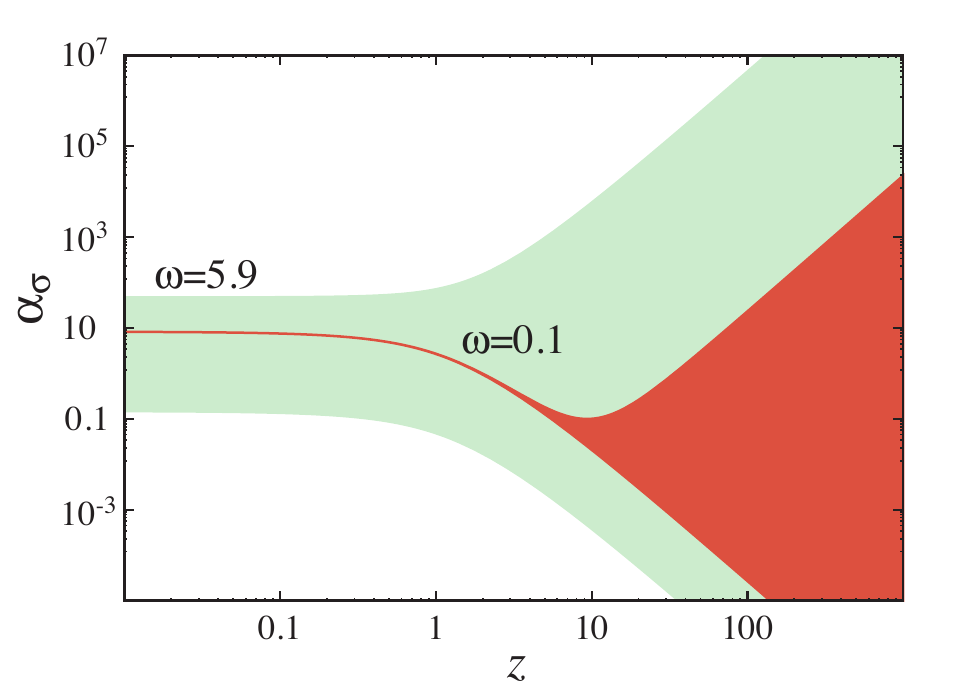}
	\caption{Evolution in redshift of the region between the upper and lower bounds upper bound  for $\alpha_\sigma$ \eqref{ngmatter} for the $\Lambda$CDM expansion history.
	A constant $\alpha_\sigma$ that satisfies the bounds exists for $\omega=5.9$ (light green shaded region) but does not for $\omega=0.1$ (dark red shaded region) yielding
	an additional constraint on $\omega$. In the latter case, all choices of $\alpha_\sigma>0$ are initially
	stable but evolve to an instability.
	Here  $\alpha_3=4$, $\alpha_4=9$, and $X=2$.	}
	\label{fig:bounds}
\end{figure}

On the other hand 
the appearance of the factor $\omega/(6-\omega)$ in \eqref{ul2} for $r$ has important
consequences for the evolution of the upper bound near $a=1$ or redshift $z=0$.   
For $\omega \rightarrow 0$, the growing part of $r$ is suppressed and 
near $z=0$ can drop below the growth of $H^2$ thereby tightening the upper bound.
Figure \ref{fig:bounds} illustrates these properties by showing the region between the
upper and lower bounds as a function of redshift $z$ for parameter set \eqref{exax} with $\omega=5.9$ (light green shaded) compared with $0.1$ (dark red shaded). 
For $\omega=5.9$, it is possible to choose a constant $\alpha_\sigma$ that satisfies the bound for all $z$ whereas it is impossible for $\omega=0.1$.
Hence, small $\omega$ is not allowed and we gain an additional constraint
beyond $0 < \omega < 6$ from requiring that the $\Lambda$CDM expansion history
be stable.

The existence of a constant $\alpha_\sigma$ which satisfy the bound \eqref{ngmatter} requires
\be
 \f{m_g^2X^2}{H_0^2}  < {\rm Min}\left[ \f{m_g^2X^2}{H(t)^2}r(t)^2 \right]
\ee
should be satisfied. By using \eqref{ul1} and \eqref{ul2}, we can rewrite this condition as
\be 
\f{\omega}{6-\omega} 
> B ,\ee
where 
\begin{equation}
B =\frac{X^2}{2(X-1)^2}
\frac{ (X-1)\alpha_3-2}{(X-1)\alpha_3-3} (\sqrt{1+\Omega_\Lambda}-1) ,
\end{equation}
 or
\begin{equation}
\frac{6 B}{1+B} < \omega < 6 .
\label{omegamatter}
\end{equation}
For instance, for parameter set \eqref{exax} and $\Omega_\Lambda=0.7$, we obtain $3.29<\omega<6$.  This means that requiring $\Lambda$CDM stability eliminates half the parameter
space that was available in the matterless case.  We emphasize that these parameters
would otherwise appear to grant stability both at the initial and current epochs.  They represent models whose fiducial metric evolve from Lorentzian to Euclidean and
back to Lorentzian.

The condition for $\alpha_\sigma$  \eqref{ngmatter}  is then
\be \label{alphamatter} \frac{X^2}{2(X-1)^2} \f{\mk{6-\omega}\Omega_\Lambda}{(X-1)\alpha_3-3}
< \alpha_\sigma < 
\f{2\omega}{(X-1)\alpha_3-2}
\kk{ 1 + \f{2\omega}{6-\omega} 
\f{(X-1)^2}{X^2}
\f{(X-1)\alpha_3-3}{(X-1)\alpha_3-2}
}. \ee
Here we used \eqref{mgvalue}.
For instance, for parameter set \eqref{exax} and $\Omega_\Lambda=0.7$, $\omega=4$, we obtain $(m_g/H_0)^2=0.7$
and $1.43<\alpha_\sigma<5$.

Thus, in addition to the conditions which we mentioned in the end of Sec.~\ref{sec-back}, we need to choose $\omega$ to satisfy \eqref{omegamatter} and $\alpha_\sigma$ to satisfy \eqref{alphamatter} given a value for $\Omega_\Lambda$ that satisfies observational
constraints on the expansion history.

\subsection{Matter with $w=1$}
\label{ssec-stiff}

Finally, for the completeness of our analysis, there is a special case that occurs
for $w=1$ or a kinetic energy dominated scalar field.  Since $3 H^2 +\dot H=0$, $r-1=0$ and (\ref{ngmatter}) would imply that no
constant $\alpha_\sigma$ can satisfy the stability bounds.   However since the derivation 
involves many expressions that assume these quantities are finite, we study this
case separately.

As in the $w\ne 1$ case, we study the conditions that would make the kinematic
matrix $K_{ij}$ for $i,j = 2,3,4$ be positive definite.
However, $\det K$ is negative definite:
\begin{align} \det K =& -\f{3m_g^4\Mpl^6X^4a^{13}[(X-1)\alpha_3-2]^2(\omega-8)^2 }
{ 64\kk{8(k/C)^2+\omega+6}^2 } \notag\\
&\times [ 64(5\omega^2-18\omega+49)(k/C)^4 +16(\omega^3+18\omega^2-61\omega+294)(k/C)^2 +(\omega^2-\omega+42)^2 ], \label{detKstiff}
\end{align}
where 
\be C \equiv m_g a X \sqrt{(X-1)\alpha_3-2}. \ee
Note that the second line of \eqref{detKstiff} is positive definite for any wavenumber under $0<\omega<6$.
Thus, there is no choice of parameters that
makes the extended quasidilaton model stable for matter with a kinetic dominated
equation of state.  This is compatible with the naive interpretation of the bound (\ref{ngmatter}).

Furthermore note that a pure $w=1$ expansion history is not strictly necessary
for the bound (\ref{ngmatter}) to have no solution.   For a multicomponent matter system,
so long as the kinetic term of the scalar field dominates the expansion, any additional subdominant
matter component with equation of state parameter $w_j<0$ will also cause a failure of solutions.   
Suppose that total energy is dominated by the kinetic term of the scalar field but has other components: $H^2=H_0^2(\Omega_s a^{-6}+\sum_j \Omega_j a^{-3(1+w_j)})$.
Then,
\be \lim_{a\rightarrow 0}\f{r-1}{H} = 
\f{\omega}{6-\omega}
\f{(X-1)^2}{X^2}
\f{(X-1)\alpha_3-3}{(X-1)\alpha_3-2}
\f{ \sum_j(1-w_j)\Omega_j a^{-3w_j} }{\Omega_\Lambda\Omega_s^{1/2}H_0}. \ee
The right hand side asymptotically vanishes as $a\to 0$ for $w_j<0$, which implies that $r/H$ approaches to $1/H$ in the past and no
constant $\alpha_\sigma$ can satisfy the stability bounds \eqref{ngmatter}. This includes the case where the additional component is from the self-accelerating background.
For $w_j=0$, solutions would only exist for special choices of parameters, e.g. $\omega \rightarrow 6$,  so that $r/H \gg 1/H$ and it allows a constant $\alpha_\sigma$ to satisfy the bound \eqref{ngmatter}.

\section{Conclusions}
\label{sec-con}

We considered  cosmological self-accelerated solutions of the extended quasidilaton theory in the presence of matter components.   By treating the matter as a scalar field with a canonical kinetic term but an arbitrary potential, we have allowed for a wide range
of background expansion histories that may occur in a cosmological setting.
Examining the quadratic Lagrangian for the scalar perturbations around these background solutions, we obtained necessary conditions for stability \eqref{ngmatter}.    While these
appear identical in form to the case without matter,  they provide time-dependent constraints on the fundamental parameters of the theory.   By  demanding the $\Lambda$CDM expansion history be stable, we obtained the constraints \eqref{omegamatter} and \eqref{alphamatter} for model parameters $\omega$ and $\alpha_\sigma$, for given value of $\Omega_\Lambda$ which are considerably stronger than the case without matter.
We also showed that the self-accelerated solution is  unstable for any choice of model parameters if the expansion is governed by matter with $w=1$ or a kinetic energy dominated scalar field.

More generally, these results arise because in this model the extended fiducial metric
is dynamical.   In particular there is nothing intrinsic to  its  dynamics that forbids an evolution
 of the fiducial metric from a Lorentzian to a Euclidean signature.   Backgrounds that evolve through
such a transition develop a ghost instability.  Thus the presence of certain types of matter can induce evolution to an instability that is not present in the initial conditions or apparent
from just the parameters of the extended quasidilaton model itself.

\acknowledgments
We thank A.\ Joyce and L.T.\ Wang for useful discussions.
HM was supported in part by JSPS Postdoctoral Fellowships for Research Abroad.
WH was supported 
by U.S.~Dept.\ of Energy
contract DE-FG02-13ER41958
and the
Kavli Institute for Cosmological Physics at the University of
Chicago through grants NSF PHY-0114422 and NSF PHY-0551142.

\appendix

\section{$\Lambda$CDM expansion history with scalar fields}
\label{app-lcdm}

In the main text, we use a scalar field to model the $\Lambda$CDM expansion history
from the matter dominated to the acceleration epoch.
It is well known that an axionic model where the field oscillates in a quadratic 
potential with $m\gg H$ satisfies these conditions averaged over oscillations.
Here we provide a novel explicit construction of an alternate case
where the field is rolling rather than oscillating and the $\Lambda$CDM expansion history
arises from attractor behavior.
We begin with the expansion history itself which can be written as
\begin{align}
a(t)&= \mk{\f{\Omega_m}{\Omega_\Lambda}}^{1/3}\sinh^{2/3}\mk{\f{3}{2}\sqrt{\Omega_\Lambda}H_0t},
\end{align}
or
\begin{align}
H(t)=\sqrt{\Omega_\Lambda} H_0 \coth\mk{\f{3}{2}\sqrt{\Omega_\Lambda} H_0t} .
\end{align}
Here we consider the $\Omega_\Lambda$ contribution to be from the quasidilaton
or more generally, a contribution that is external to the scalar field system.
Therefore, for the scalar system to combine with $\Omega_\Lambda$ to form the
$\Lambda$CDM expansion history, we require the energy in the scalar field
to scale as $\rho \propto a^{-3}$ and the pressure $p=0$.   This condition then requires
$\dot{\bar\xi}^2/2 = V(\bar\xi) = \rho/2$ or
\begin{align}
\bar\xi&=\pm \f{2\tMpl}{\sqrt{3}} \log\kk{\f{ \tanh\mk{\f{3}{4}\sqrt{\Omega_\Lambda} H_0t} }{\f{3}{4}\sqrt{\Omega_\Lambda} }},
\label{LCDMsolution}
\end{align}
and finally
\begin{align}
V&=\f{3\Omega_\Lambda H_0^2\tMpl^2}{8} \kk{ \f{3}{4}\sqrt{\Omega_\Lambda} e^{\f{\sqrt{3} \xi}{2\tMpl }}  - \mk{\f{3}{4}\sqrt{\Omega_\Lambda} e^{\f{\sqrt{3} \xi}{2\Mpl }} }^{-1} }^2.
\end{align}
This solution (\ref{LCDMsolution}) is an attractor of this potential.  It  also holds for a true cosmological constant rather than the quasidilaton effective cosmological constant with the replacement $\tMpl \rightarrow \Mpl$. 

Note that this system differs from axionic scalar field solutions that also satisfy
the $\Lambda$CDM expansion history.   In our solution the kinetic and potential energies
are set to be equal instantaneously whereas for an axion they only average to the same
values over many oscillations.   This difference also appears in the dynamics of 
perturbations.   Here they are gravitationally stable due to the field fluctuations having
sound speed unity in a slowly varying background.   In the axion case, the rapid 
oscillation of the background allows for growing modes in the energy density perturbations
that behave like CDM.    Unlike the axionic case, this model is an example of
a system that is indistinguishable from $\Lambda$CDM from the expansion history but
easily distinguishable in the growth of structure.

\section{Eigensystem for sub-kinematic matrix}
\label{app-es}

Here we give explicit forms for the  eigenvalues and eigenvectors of the sub-kinematic matrix for $(\Psi, \d\sigma, \d\xi)$ used in Sec. \ref{ssec-matter}.
Diagonalizing the matrix yields the eigenvalues
\be \lambda_1\equiv 0,\quad \lambda_2\equiv \f{1}{4}(p_0- \sqrt{q}), \quad \lambda_3\equiv \f{1}{4}(p_0+ \sqrt{q}), \ee
and the corresponding eigenvectors 
\be \vec{v}_1^T\equiv (1/\Xi, 1/\Xi, 1), \quad \vec{v}_2^T\equiv(v_{21}, v_{22},1), \quad \vec{v}_3^T\equiv(v_{31}, v_{32}, 1), \ee
where recall $\Xi= {\dot {\bar\xi}}/{\Mpl H}$ and
\begin{align}
v_{21} &\equiv \f{p_1-(\omega+\Xi^2)\sqrt{q}}{d+ \Xi\sqrt{q}}, \quad 
v_{22} \equiv \f{p_2+ \omega\sqrt{q}}{d+ \Xi\sqrt{q}}, \nonumber\\
v_{31} &\equiv \f{p_1+ (\omega+\Xi^2)\sqrt{q}}{d- \Xi\sqrt{q}}, \quad 
v_{32} \equiv \f{p_2- \omega\sqrt{q}}{d- \Xi\sqrt{q}},
\end{align}
and 
\begin{align}
q\equiv& 16(6-\omega)^2(r^2-1)^2[\Xi^4+2\Xi^2+(2\omega-1)^2] \mk{\f{k}{aH}}^4\notag\\
&-8\omega(6-\omega)(r^2-1)(\Xi^2+\omega-6)[(\omega+6)\Xi^4+2(\omega+6)\Xi^2+(2\omega-1)(13\omega-6)] \mk{\f{k}{aH}}^2\notag\\
&+\omega^2(\Xi^2+\omega-6)^2[(\omega+6)^2\Xi^4+2(\omega+6)^2\Xi^2+(13\omega-6)^2], \nonumber\\
d\equiv& -4(6-\omega)(r^2-1)\Xi(\Xi^2+1)\mk{\f{k}{aH}}^2
+\omega\Xi(\Xi^2+\omega-6)[(\omega+6)\Xi^2+2\omega^2-\omega+6], \nonumber\\
p_0\equiv& 4(\omega-6)(r^2-1)(\Xi^2+2\omega+1)\mk{\f{k}{aH}}^2
+\omega(\Xi^2+\omega-6)[(\omega-6)\Xi^2-(11\omega+6)],\nonumber\\
p_1\equiv& 4(6-\omega)(r^2-1)[\Xi^4+(\omega+1)\Xi^2+\omega(2\omega-1)]\mk{\f{k}{aH}}^2\notag\\
&
-\omega(\Xi^2+\omega-6)[(\omega+6)\Xi^4+(\omega+1)(\omega+6)\Xi^2+\omega(13\omega-6)], \nonumber\\
p_2\equiv& -4\omega(6-\omega)(r^2-1)[\Xi^2+2\omega-1]\mk{\f{k}{aH}}^2
+\omega^2(\Xi^2+\omega-6)[(8-\omega)\Xi^2-13\omega+6]. 
\end{align}
The following relations also help simplify the derivation:
\begin{align} 
\Xi^2+\omega-6&=-\f{2}{H^2}\mk{ \Lambda+\Lambda_X+\f{V}{\Mpl^2} } \notag\\
&=-2\mk{1-\f{\omega}{6}}\f{3H^2+\dot H}{H^2},
\end{align}
from the background equations and
\begin{align}
v_{21}+v_{22}+\Xi &=0,\nonumber\\
v_{31}+v_{32}+\Xi &=0,\nonumber\\
v_{21}v_{31}+v_{22}v_{22}+1 &=0,
\end{align}
from the orthogonality of the eigenvectors.

\bibliography{refs}
\end{document}